\documentclass[]{spie}  %>>> use for US letter paper
\usepackage{amsmath,amsfonts,amssymb}
\usepackage{float}
\usepackage{graphicx}
\usepackage[colorlinks=true, allcolors=blue]{hyperref}
\usepackage{multirow}
\usepackage{caption}

\title{Simulating the efficacy of the implicit-electric-field-conjugation algorithm for the Roman Coronagraph with noise}

\author[a]{Kian Milani}
\author[b]{Ewan Douglas}
\author[b]{Sebastiaan Haffert}
\author[b]{Kyle Van Gorkom}
\affil[a]{James C. Wyant College of Optical Sciences}
\affil[b]{Steward Observatory}

\begin{document}
\maketitle


\begin{abstract}
The Roman Coronagraph is expected to perform its high-order wavefront sensing and control (HOWFSC) with a ground-in-the-loop scheme due to the computational complexity of the Electric-Field-Conjugation (EFC) algorithm. This scheme provides the flexibility to alter the HOWFSC algorithm for given science objectives. A new alternative implicit-EFC algorithm is of particular interest as it requires no optical model to create a dark-hole, making the final contrast independent of the model accuracy. The intended HOWFSC scheme involves running EFC while observing a bright star such as $\zeta$ Puppis to create the initial dark-hole, then slew to the science target while maintaining the contrast with low-order WFSC over the given observation. 

Given a similar scheme, the efficacy of iEFC is simulated for two coronagraph modes, namely the Hybrid Lyot Coronagraph (HLC) and the wide-field-of-view Shaped-Pupil-Coronagraph (SPC-WFOV). End-to-end physical optics models for each mode serve as the tool for the simulations. Initial monochromatic simulations are presented and compared with monochromatic EFC results obtained with the FALCO software. Various sets of calibration modes are tested to understand the optimal modes to use when generating an iEFC response matrix. Further iEFC simulations are performed using broadband images with the assumption that $\zeta$ Puppis is the stellar object being observed. Shot noise, read noise, and dark current are included in the broadband simulations to determine if iEFC could be a suitable alternative to EFC for the Roman Coronagraph.
\end{abstract}

\section{Introduction}
\label{sec:intro}
With thousands of exoplanets having been discovered primarily via indirect detection methods including transit photometry and radial velocity measurements, further study of exoplanets is desired to understand the dynamics of planet formation, interactions with debris disks, and the potential for an exoplanet to be habitable. These studies can be enabled if the challenges of direct-imaging are overcome, of which the two primary challenges are the relatively small angular separation between the host star and the exoplanet as well as the high planet to star flux ratio. For perspective, a Sun-Earth analog at 10 parsecs will have an angular separation of about 0.1 arcsec and a flux ratio of about $10^{-10}$ (meaning the planet is dimmer than the star by a factor of ~10 billion). While many coronagraphs have been demonstrated on ground-based telescopes and contrast levels of ${\sim}10^{-6}$ have been achieved, atmospheric turbulence and absorption in certain bandpasses limit the performance of these ground based instruments. A key milestone for direct-imaging will be the launch of the Nancy Grace Roman Space Telescope, which will carry an onboard coronagraph with multiple modes for imaging and spectral characterization\cite{riggs-flight-mask-designs}. 

The Roman Coronagraph will serve as a demonstration of technologies deemed crucial for the detection of Earth-like exoplanets for future spaceborne coronagraphs. A few of the key technologies to be demonstrated are high-order wavefront sensing and control (HOWFSC) with two large deformable mirrors, low-noise photon-counting detectors, and two distinct coronagraph concepts\cite{riggs-flight-mask-designs}. In concert with the demonstration of these technologies, the Coronagraph will bridge the gap between current direct-imaging capabilities and those desired for future telescopes to image Earth-like planets at $10^{-10}$ contrast levels. 

While the requirement for the Coronagraph is to achieve $10^{-7}$ contrast levels, current performance estimates attained with models of the Coronagraph indicate raw contrasts of $10^{-8}$ to $10^{-9}$ will be achieved\cite{riggs-flight-mask-designs}. A crucial step in achieving these contrasts will be the use of Electric-Field-Conjugation (EFC) algorithm, where a high-contrast region designated as the dark-hole is created through sequential estimation and minimization of the electric field intensity. While EFC is demonstrated to achieve high-contrasts on many testbeds, EFC requires an optical model to perform an initial estimation of the electric field at the focal plane of the coronagraph. As such, EFC has been demonstrated to be sensitive to model errors, particularly those associated with errors in the model of the deformable mirror\cite{potier-comparing-fpwfs}. A completely data driven focal plane wavefront control strategy such as implicit-EFC could be a desirable alternative to EFC if calibrating the EFC model proves to be a greater challenge when the Coronagraph is in orbit.  

Presented in this paper are HOWFC results obtained by simulating both EFC and iEFC. For EFC simulations, the FALCO software\cite{sidick-falco-2} is used, in turn utilizing the roman\_phasec\_proper models of the Coronagraph with PROPER\cite{krist-proper} acting as the backend propagator. For iEFC simulations, similar end-to-end optical models utilizing Fresnel propagation and surface errors of individual optics are used, however, the models are built utilizing POPPY\cite{douglas-poppy-accelareted-modeling} as the backend propagator. The agreement between the PROPER and POPPY models has been demonstrated in previous work\cite{milani-updated-roman-models} and the simulations here present the current agreement with the updated model versions as well. The primary motivation for using the POPPY models is to accelerate computations as POPPY now includes CuPy as an option for computations, enabling end-to-end propagation entirely on a GPU. Using both FALCO and the POPPY models, initial monochromatic simulations of both EFC and iEFC are presented for the HLC and SPC-WFOV imaging modes of the Coronagraph. Further broadband simulations are performed for iEFC with shot noise, read noise, and dark current included to estimate the feasibility and calibration times that would be required for iEFC. While two DMs are utilized for both imaging modes, this investigation has primarily been focused on creating half dark-holes for the HLC and a full annular dark-hole for the SPC-WFOV mode. The motivations for this are expanded upon in Sections \ref{sec:hlc-mono-sims} and \ref{sec:spc-wfov-mono-sims}. The POPPY models\cite{milani-cgi-phasec-poppy} and iEFC algorithms\cite{milani-roman-cgi-iefc} used for these simulations are available in public repositories\footnote{\href{https://github.com/kian1377/cgi_phasec_poppy}{https://github.com/kian1377/cgi\_phasec\_poppy}, \href{https://github.com/kian1377/roman-cgi-iefc}{https://github.com/kian1377/roman-cgi-iefc}}. The PROPER models and FALCO software suite are available through separate distribution channels, but are nonetheless publicly available.

\section{Definitions and algorithms}
The metric used to evaluate the contrast attained by each algorithm is normalized intensity, defined as $$I_{image}(x,y)/max(I_{uo}(x,y))$$ where $I_{image}(x,y)$ is the intensity of a raw coronagraphic image and $I_{uo}(x,y)$ is the intensity of the unocculted image (no coronagraphic masks in the optical system). This is the same metric used in FALCO\cite{sidick-falco-2} and while this does not directly evaluate contrast due to the field dependent intensity of off-axis companions, it is a reasonable metric to evaluate the coronagraph performance. In the following sections, the terms contrast and normalized intensity will be used interchangeably using the definition provided here. 

\subsection{EFC}
The fundamental concept of EFC is that the irradiance in the image plane can be minimized by computing DM commands that generate an electric-field in the focal plane that will destructively interfere with the measured electric-field\cite{give'on-bb-wavefront-correction}. This is done by using a model of the instrument to calculate a Jacobian matrix relating the response of the focal plane electric-field field to the actuators on a DM. A control matrix is generated from the Jacobian by inverting with a given regularization method. The matrix-vector product between the control matrix and the focal-plane electric-field will yield the actuator commands that result in that electric-field. By applying the negative of the calculated actuator commands, the electric-field can be destructively interfered. A critical component to EFC is how the electric-field in the focal plane is estimated prior to the actuator command being calculated. A common technique to perform this estimation is the use of Pairwise-Probing (PWP), where a set of DM probes are used to generate phase diversity within the focal plane. Measurements taken with each probe can then be used to calculate the electric-field with the use of the instrument model. 

FALCO is a software suite designed to perform EFC in a computationally efficient manner with a unique method for calculating the response of the instrument actuators using a compact (Fraunhofer) model of the instrument\cite{riggs-falco-1}. This enables fast computations of the Jacobian as well as relinearization of the Jacobian during iterations of EFC. For a given simulation of EFC, there are many model parameters that can be adjusted within FALCO. The simulations performed here are configured very similarly to the examples of the Roman Coronagraph modes provided in the FALCO software suite. For most simulations, the ideal case of EFC was considered by assuming that perfect knowledge of the electric-field could be attained. In simulations, this uses the end-to-end model to directly calculate the electric-field at the focal plane. An example of EFC with PWP is also provided for the HLC mode in Section \ref{sec:hlc-mono-falco-sims} to demonstrate a more realistic scenario. 

\subsection{iEFC}
Unlike EFC, iEFC includes no explicit calculation of the electric-field between single iterations of the algorithm, hence the name implicit-EFC. Instead, iEFC transforms difference images directly into DM commands using a single response matrix as described in Haffert et al\cite{haffert-iefc}. This response matrix relates DM modes to modulated difference images. A key distinction between EFC and iEFC is that with the electric-field explicitly calculated in EFC, the focal plane speckles are destructively interfered by injecting the opposite electric-field with the calculated DM shape(s). Since the response matrix for iEFC is generated purely with modulated difference images, destructively interfering the speckles is not directly possible with iEFC. Instead, if the difference images themselves are minimized, the electric-field also becomes minimized since they are a linear proxy for the electric-field\cite{haffert-iefc}. Further description of the iEFC fundamentals can be found in Haffert et al\cite{haffert-iefc}, where EFC and iEFC were demonstrated to have comparable performance with simulations of an scalar vortex coronagraph (SVC). For clarity, the diagram below illustrates the steps required to calibrate iEFC for a set of modes on a single DM. 

\begin{figure}[H]
    \centering
    \includegraphics[scale=0.3]{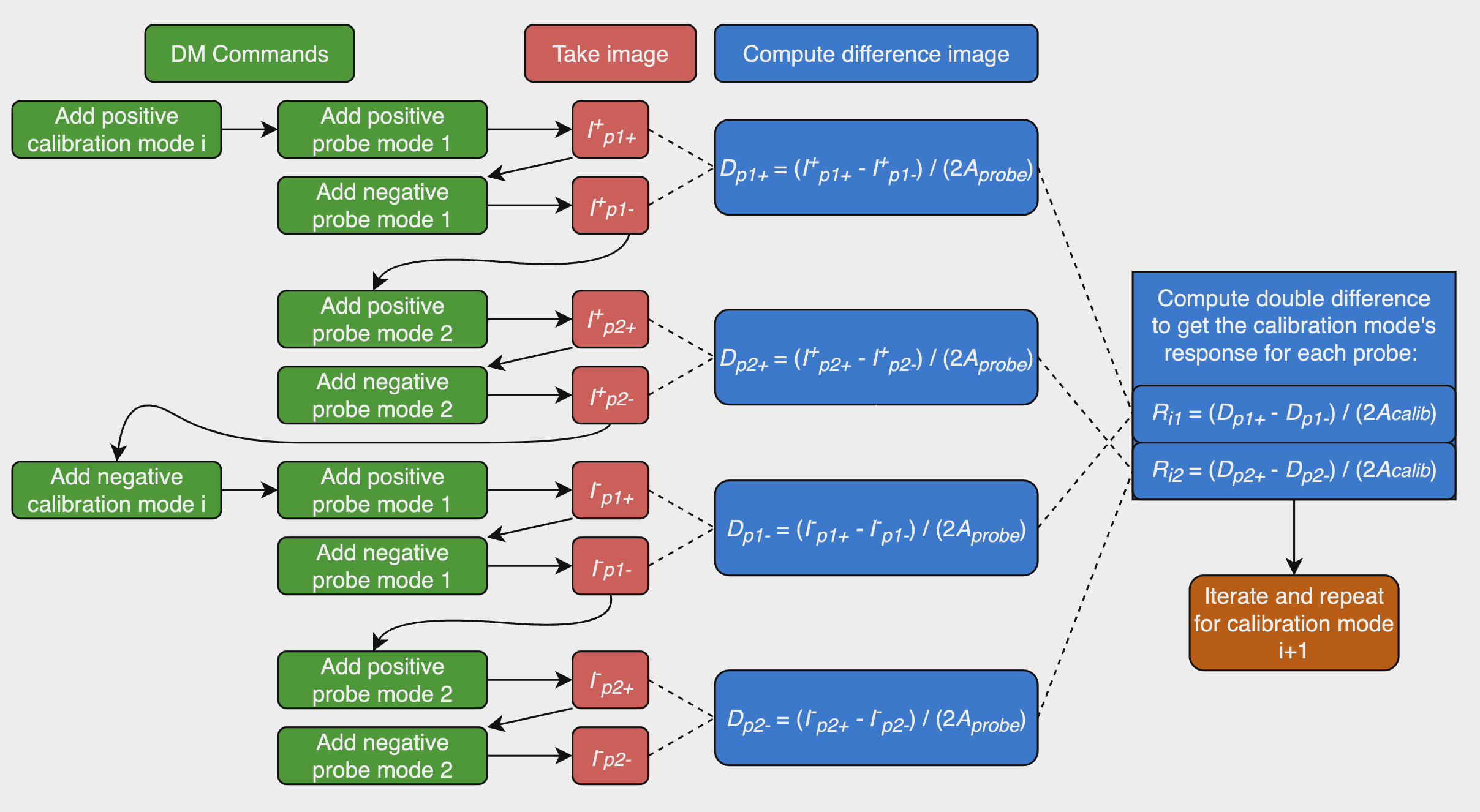}
    \caption{This diagram illustrates the critical steps for generating a response matrix for iEFC by calibrating a chosen set of DM modes. Note that in this diagram, only 2 probe modes are being utilized to generate the difference images for a calibration mode, but more probes could be used if desired.}
    \label{fig:iefc-calibration}
\end{figure}

% A detailed description of all the variables in the figures can be found Table \ref{tab:} in the appendix. 

\begin{figure}[H]
    \centering
    \includegraphics[scale=0.3]{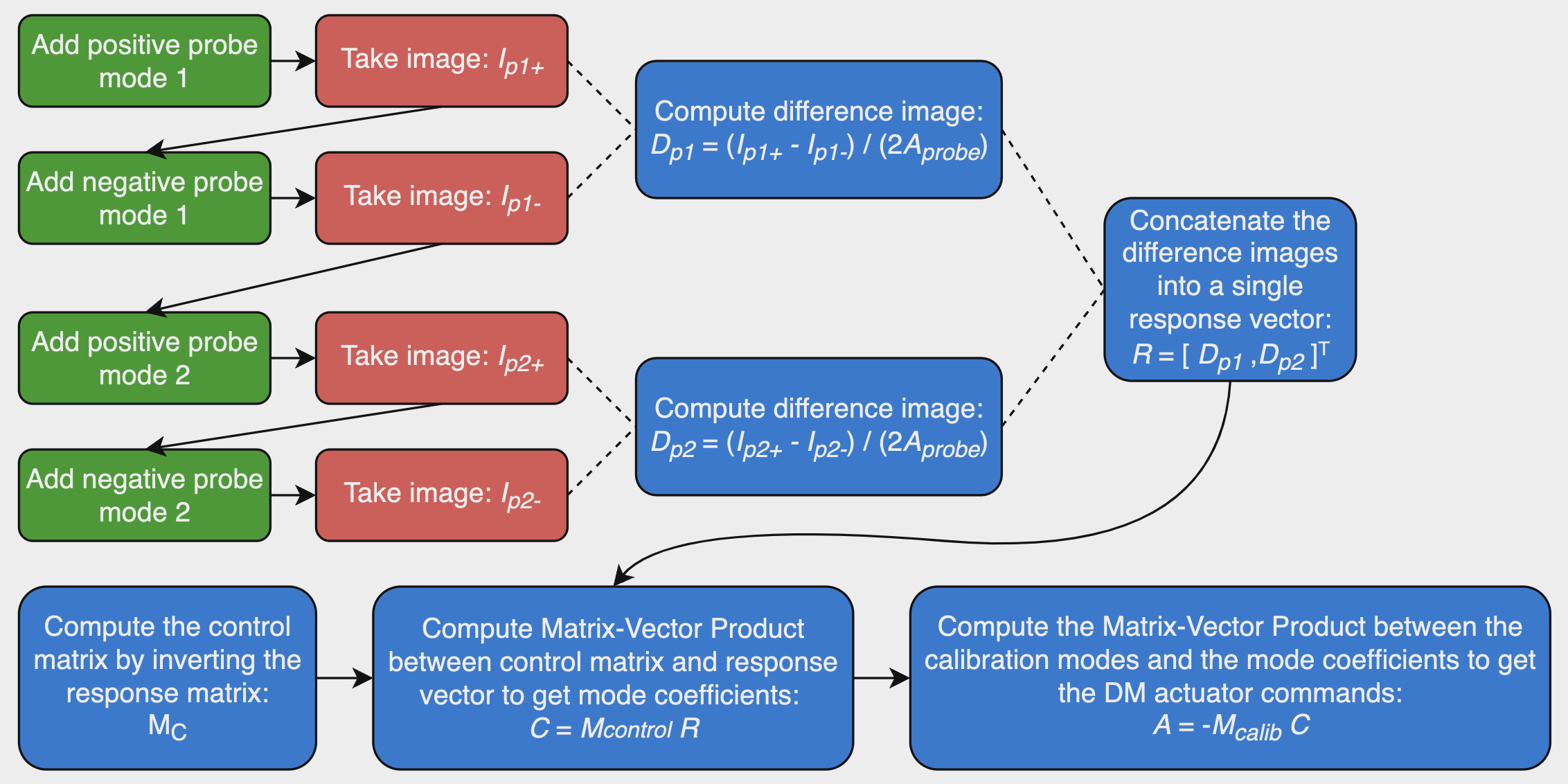}
    \caption{Flowchart illustrating the steps to run a single iteration of iEFC.}
    \label{fig:iefc-algorithm}
\end{figure}

Assuming two probes are used, the final response matrix must have array dimensions of $2N_{pix} \times N_{modes}$ where $N_{pix}$ is the number of pixels within the desired control region of the focal plane and $N_{modes}$ is the total number of calibration modes used. In general, if more than two probes are used then the shape of the final response matrix should be $N_{probes} N_{pix} \times N_{modes}$. Note that in the calibration diagram above, 8 images are required to calibrate a single mode and using more probe modes will increase the number of images required. Here, the iEFC controller is extended to utilize two deformable mirrors for application to the Roman Coronagraph by calibrating modes on both DMs while using the first DM for all the probes. Different sets of modes could be utilized for the two DMs, but for simplicity, the simulations here utilize the same modes for each DM. With two DMs, the number of modes to calibrate increases by a factor of 2 (assuming the same modes are used for each DM). In practice, the response matrices for the modes on each DM are concatenated to produce 2 DM response matrix with array dimensions of $N_{probes} N_{pix} \times N_{modes}$. Figure \ref{fig:iefc-algorithm} now demonstrates how to run iEFC with a single DM once all modes have been calibrated. 

% Running iEFC with two DMs remains very similar except that the matrix-vector product between the calibration modes matrix $M_{calib}$ and the modal coefficients vector $C$ must be performed twice as demonstrated in the equations below. Here, $C$ contains the the modal coefficients for the modes on DM1 and DM2 respectively while $A_{DM2}$ and $A_{DM2}$ are the actuator commands for each DM. If DM modes are calibrated are calibrated in terms of actuator voltages, then  $A_{DM}$ will be calculated in volts. For these simulations, all DM modes are calibrated and calculated directly in terms of actuator heights. 

Running iEFC with two DMs remains very similar except that the matrix-vector product between the calibration modes matrix $M_{calib}$ and the modal coefficients vector $C$ now yields the concatenated actuator commands for each DM as demonstrated by the equation $$A = -M_{calib} C = [A_{DM1}, A_{DM2}].$$ Here, $C$ will have $N_{modes}$ elements (although the number of modes is twice that of the the single DM scenario) while the actuator command vector will have $2N_{act}^2$ elements which can be decomposed into the commands for each DM. If DM modes are calibrated in terms of actuator voltages, then  $A$ will be calculated in volts. For these simulations, all DM modes are calibrated and calculated directly in terms of actuator heights. 

The two DM iEFC controller is initially tested using an HCIpy\cite{por-hcipy-2018} scalar vortex coronagraph model to verify the capability of creating an annular dark-hole. The model uses a charge 6 vortex with a 90\% diameter Lyot stop. The wavelength for the simulation is 650nm with two $34\times34$ actuator DMs. The DM pupil is 10mm in diameter with 1m separation between the DMs. The DM modes used for calibration are Fourier modes, which are a set of sinusoidal shapes chosen to generate strong responses in the desired control region. The probes used are single actuator pokes of two neighboring actuators. The result of the dark-hole is presented below in units of normalized intensity. 

\begin{figure}[H]
    \centering
    \includegraphics[scale=0.5]{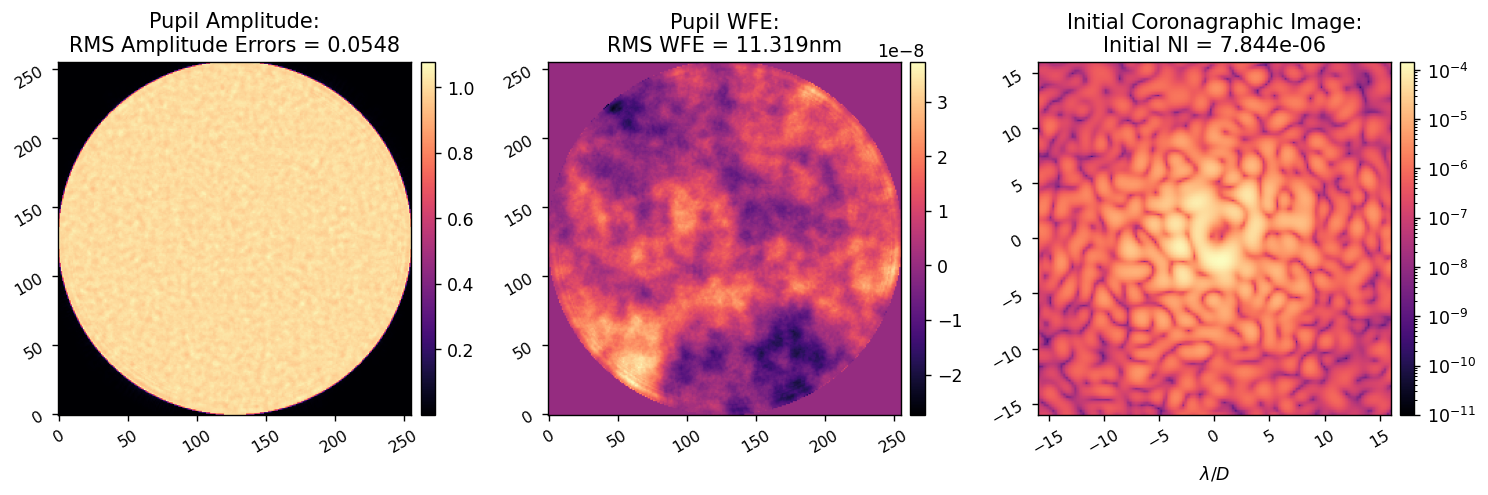}
    \includegraphics[scale=0.5]{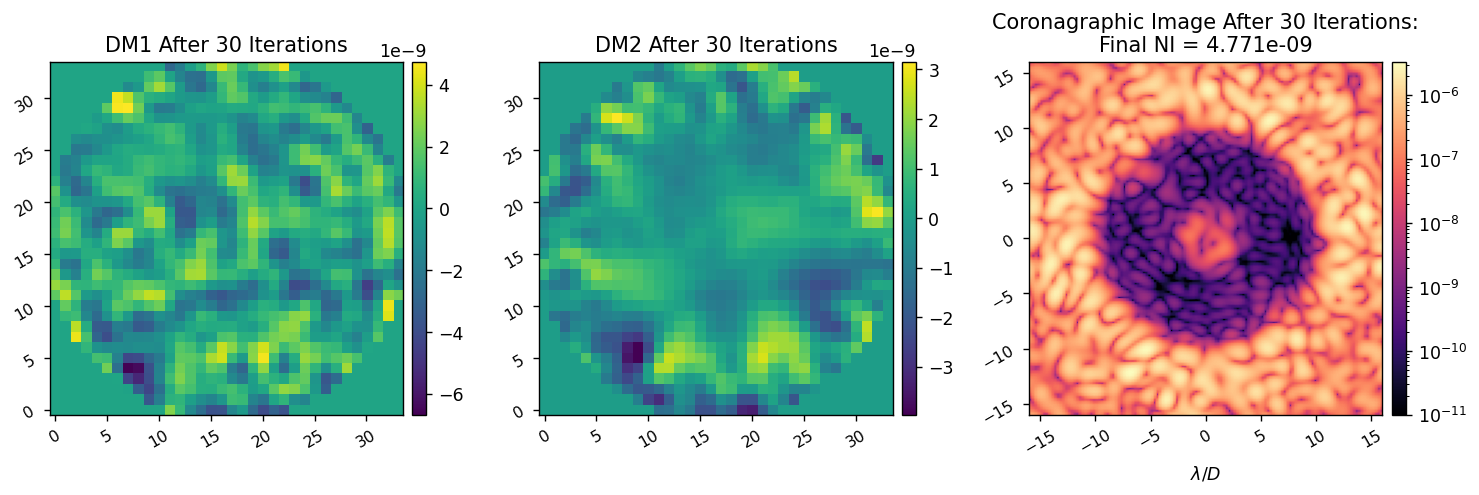}
    \caption{Annular dark-hole created using iEFC with a vortex coronagraph model. The specified ROI for the dark-hole extends from an IWA of 3$\lambda/D$ to $10\lambda/D$. Th final mean contrast is measured to be ~1.6E-9 after 30 iterations.}
    \label{fig:iefc-2dm-annular-dark-hole}
\end{figure}

While the result above is for an ideal monochromatic vortex coronagraph, the application of iEFC with 2 DMs is clearly found to be suitable for annular dark-holes. This same algorithm is now applied to both the HLC and SPC-WFOV modes using the physical optics models described in Section \ref{sec:intro}. 

\section{HLC Monochromatic Simulations}

\label{sec:hlc-mono-sims}
For this imaging mode, both EFC and iEFC were applied to create only a half dark-hole while still utilizing both DMs. This is largely due to the difficulty of creating a full dark hole because of the frequency with which relinearization of the response matrix or Jacobian is found to be required for HOWFSC to be successful with this mode. With FALCO, relinearization is performed computationally using the compact (Fraunhofer) model of the system. The response to each actuator in the control region is remeasured at specified iterations using FALCO's unique algorithm for estimating the response using sub-sampled regions of the pupil\cite{riggs-falco-1}. For iEFC, the data driven nature of the algorithm's calibration requires generating the entire response matrix again using instrument images. Note that in practice, this recalibration can become cumbersome if long exposure times are required, while in simulations, recalibrating is a computationally expensive process that makes the simulations much more time-consuming using end-to-end models. Recent work from Gersh-Range et al.\cite{gersh-range-pupil-error-spc-bowtie} on the SPC spectroscopy/bowtie Coronagraph mode has demonstrated that using two DMs to create a half dark-hole can be beneficial as each DM requires less stroke and the DMs can more effectively correct over a large band. These results also motivated the application of a half dark-hole as using less stroke with the DMs could reduce the necessity for relinearization. While the requirement for the HLC is to achieve $10^{-7}$ contrast for an annular dark-hole, a half-dark hole could still yield many scientific results for direct-imaging. 

\begin{figure}[H]
    \centering
    \raisebox{-0.5\height}{\includegraphics[scale=0.4]{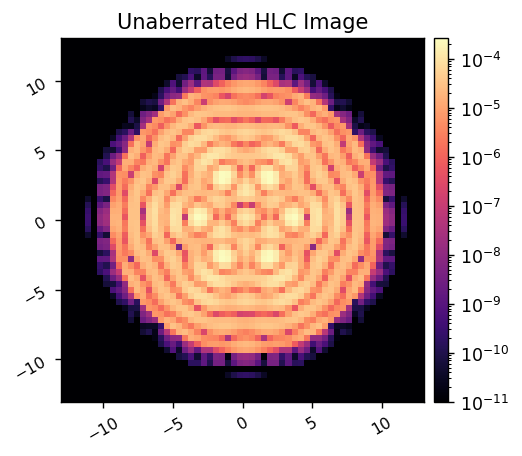}}
    \raisebox{-0.5\height}{\includegraphics[scale=0.4]{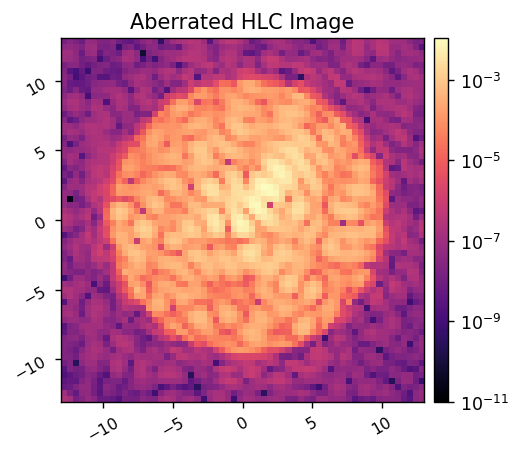}}
    \raisebox{-0.5\height}{\includegraphics[scale=0.35]{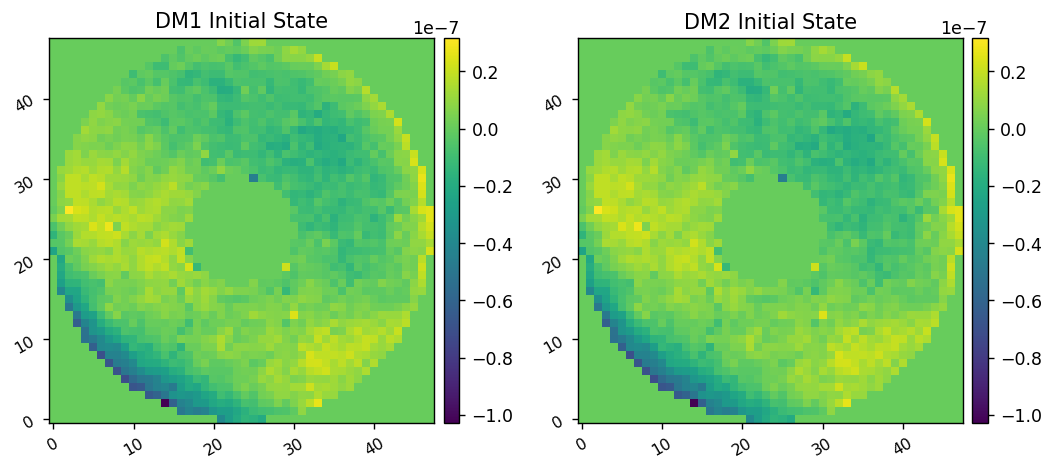}}
    \raisebox{-0.5\height}{\includegraphics[scale=0.4]{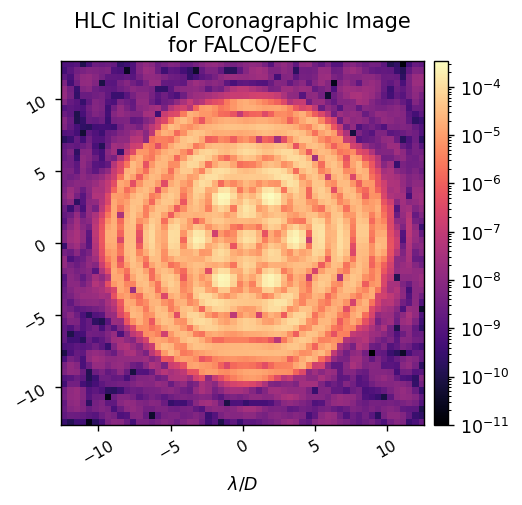}}
    \raisebox{-0.5\height}{\includegraphics[scale=0.4]{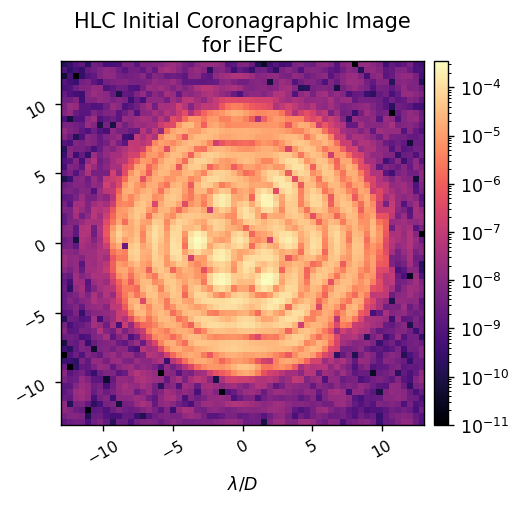}}
    \caption{The starting state of the HLC mode for both FALCO/EFC and POPPY/iEFC simulations. The initial DM commands utilized are the commands to flatten the wavefront in the optical system to nominally restore the coronagraph to an unaberrated state. These DM commands are included in the distribution of the roman\_phasec\_proper package. }
    \label{fig:hlc-initial-states}
\end{figure}

Presented in Figure \ref{fig:hlc-initial-states} are the initial states for the HLC DMs and the normalized intensity images of each model. The purpose of the initial DM shapes chosen is to correct WFEs in the HLC mode, but not to create a dark-hole. Note that because different end-to-end models were used for FALCO and iEFC simulations, there is a slight difference between the morphology of the initial coronagraphic images between the FALCO and iEFC models. Specifically, the image from the FALCO model is slightly less aberrated, likely due to small discrepancies between the DM models used in PROPER and POPPY. These differences are not significant for the purposes of demonstrating iEFC since the mean contrasts within the control region of the iEFC and FALCO models are 3.178E-05 and 3.00E-5 respectively.

\subsection{FALCO}
\label{sec:hlc-mono-falco-sims}
\begin{figure}[H]
    \centering
    \raisebox{-0.5\height}{\includegraphics[scale=0.25]{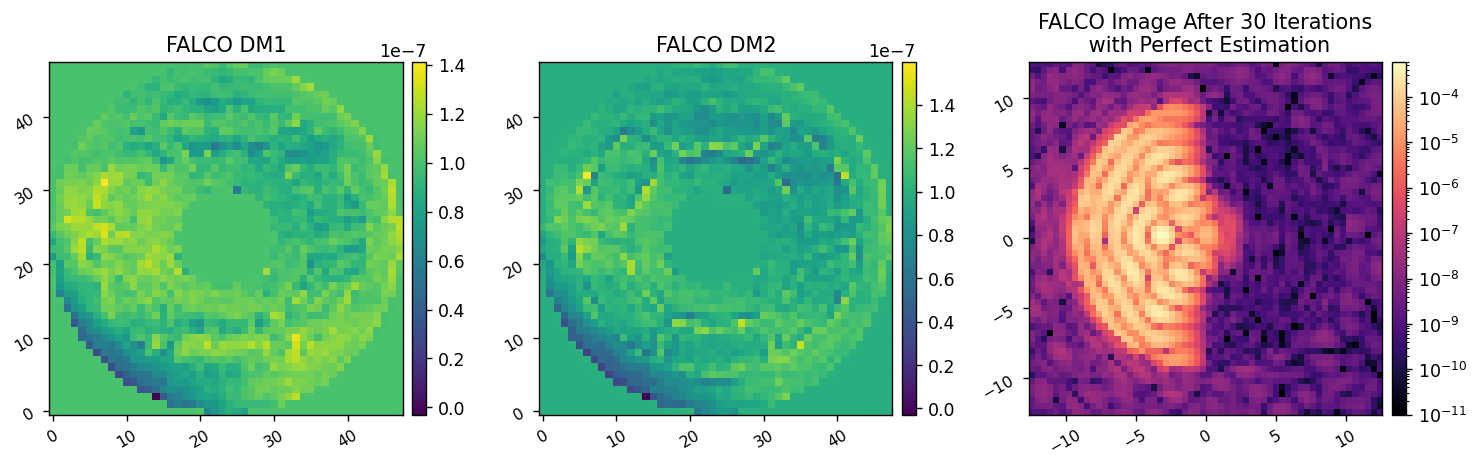}}
    % \raisebox{-0.5\height}{\includegraphics[scale=0.3]{figures/hlc_falco_contrast_per_iter.png}}
    \caption{After running 30 iterations of EFC with perfect estimation of the electric-field, the final contrast within the control region is 1.77E-9.}
    \label{fig:hlc-falco-results}
\end{figure}

Since the goal for this mode is to create a half dark-hole, the Jacobian is relinearized every 5 iterations since rather than every iteration given less stroke is required from each DM. Figure \ref{fig:hlc-falco-results} presents the dark-hole achieved when assuming a perfect estimation of the electric-field after 30 iterations with 6 Jacobian computations. The RMS of each DM command required to obtain the dark-hole presented are 6.1nm and 8.6nm respectively.

\begin{figure}[H]
    \centering
    \includegraphics[scale=0.5]{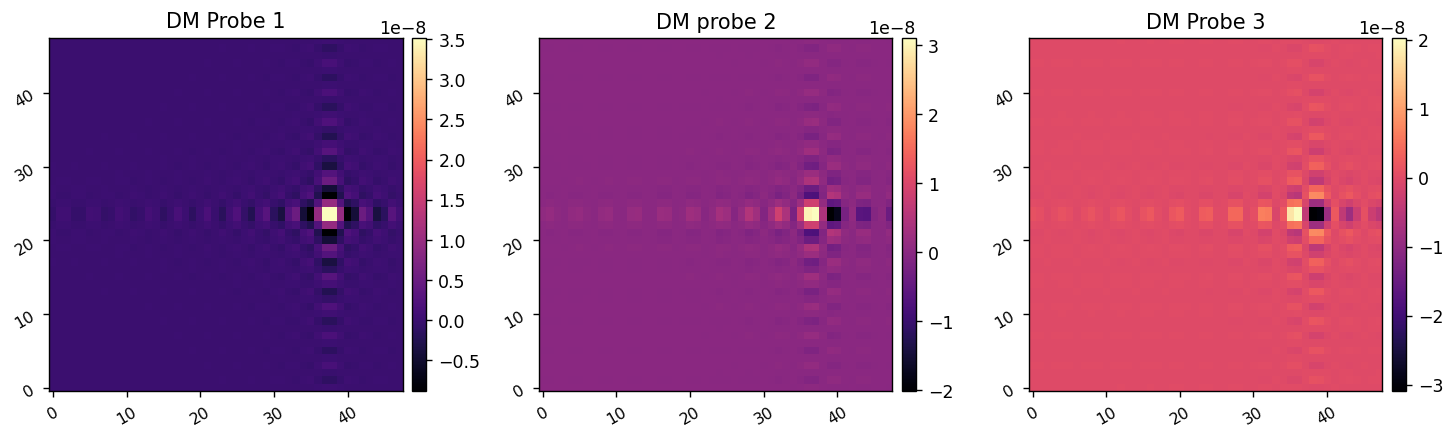}
    \caption{Example DM probes used during pairwise probing for E-field estimation. These probes are a combination of a sinc and a sinusoid specifically calculated to generate a strong response within desired region of interest. The examples shown were calculated for the HLC mode, where the region of interest extends to about $10\lambda/D$. The probes are shifted to avoid the obscured region of the pupil where actuators have much weaker response.}
    \label{fig:falco-probes}
\end{figure}

While very deep contrasts can be achieved in a short number of iterations assuming that the electric-field is calculated perfectly, PWP greatly affects the performance of EFC. Figure \ref{fig:falco-probes} illustrates the three pairs of probes used to estimate the electric-field with PWP. The batch-process PWP algorithm within FALCO was specified for the estimation. 

\begin{figure}[H]
    \centering
    \raisebox{-0.5\height}{\includegraphics[scale=0.45]{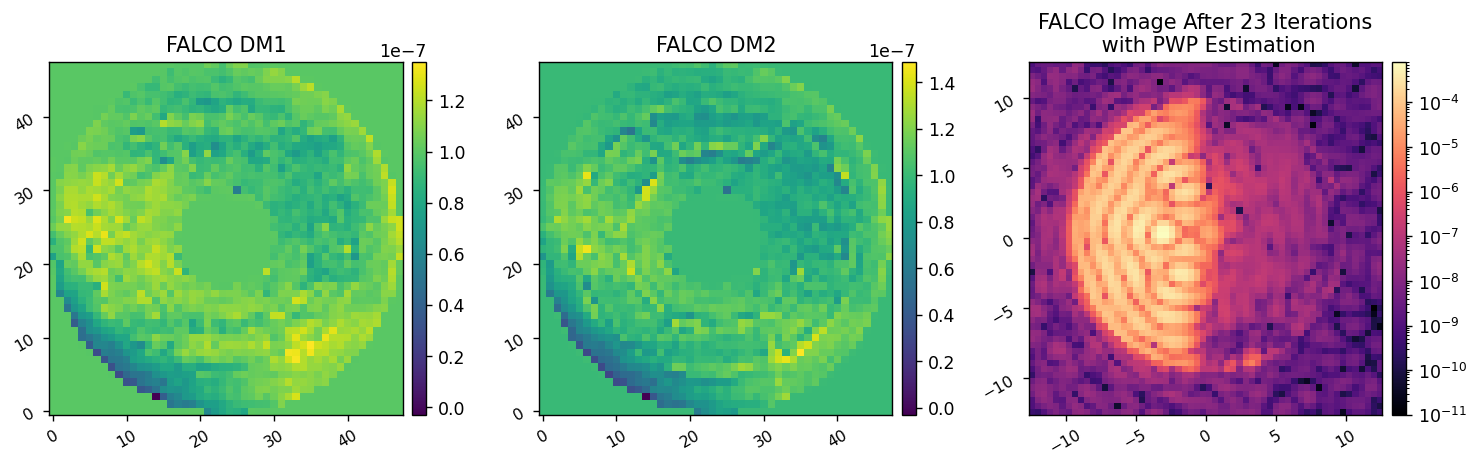}}
    \caption{The DM commands and the final image from FALCO's simulations of EFC with PWP. The mean contrast within the half dark-hole is 1.14E-7.}
    \label{fig:hlc-falco-pwp-results}
\end{figure}

The final result with PWP implemented is presented in Figure \ref{fig:hlc-falco-pwp-results}. Here, EFC does not converge to a dark-hole as effectively since the estimation of the electric-field is critical for the actuator commands to produce destructive interference. This demonstrates why a non-model based algorithm is desirable given that the model is also a critical tool for PWP to be functional. Here, the final EFC result is the 23rd iteration because the EFC solutions began to diverge after this. A better choice of probes and probe amplitude could solve this, however, this is a simple demonstration of how PWP can degrade the performance of EFC. 

\subsection{iEFC}
\begin{figure}[H]
    \centering
    \includegraphics[scale=0.25]{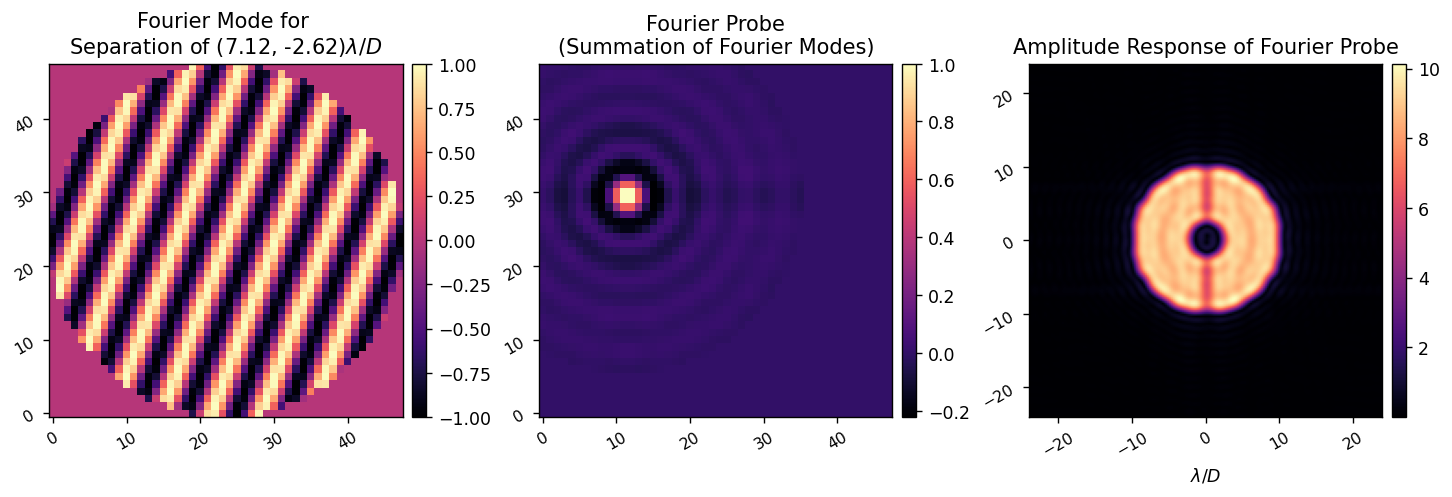}
    \caption{From left to right are an example a single Fourier mode, a Fourier probe, and the focal plane response of the Fourier probe used to generate the response matrix for the HLC. The Fourier probe is the sum of all cosine Fourier modes calculated to span the desired control region of the focal plane. Two Fourier probes are used, one being the sum of the cosine modes and the other being the sum of the sine modes. Note that the Fourier probe is shifted such that it is within the region of the DM where the actuators are not obstructed by the telescope pupil or the Lyot stop. The Fourier Transform of this probe displayed on the right illustrates that it generates a uniform response in the desired control region. }
    \label{fig:mode-examples}
\end{figure}

Initial iEFC simulations with the HLC mode proved difficult to yield reasonable contrast results. Three sets of calibration modes were tested including Fourier modes, Hadamard modes, and single actuator pokes. Likely due to the geometry of the Roman pupil, single actuator pokes were found to be the most optimal. The Fourier modes and Hadamard modes would span all actuators on the DMs, but the combination of the pupil and the Lyot stop would strongly influence which actuators have a significant response in the focal plane. The probe modes used were a sum of cosine and sine Fourier modes generated to span the desired control region. Figure \ref{fig:mode-examples} presents an example of the Fourier probes used for calibration along with the theoretical focal plane response of the probe. 

\begin{figure}[H]
    \centering
    \includegraphics[scale=0.25]{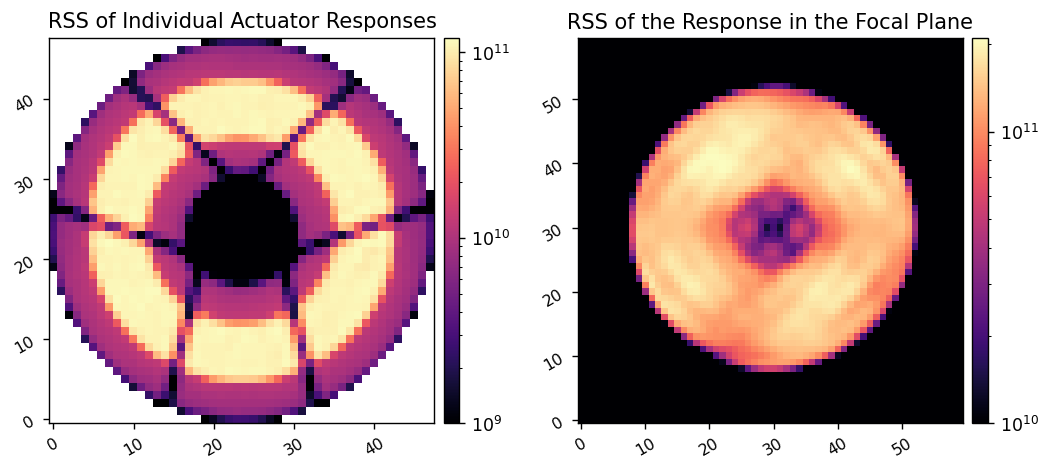}
    \caption{To the left is the RSS of the response of each actuator modes calibrated for iEFC and to the right is the RSS of the response for each pixel in the focal plane. Together, these demonstrate the actuators that have a strong influence in the focal plane and hence the actuators that will be heavily used during iEFC as well as the region in the focal plane that was well calibrated.}
    \label{fig:hlc-response}
\end{figure}

After calibration is completed, the quality of the calibration can be visualized based on the root-sum-squared (RSS) of the responses for each actuator mode and for each pixel in the focal plane. Figure \ref{fig:hlc-response} presents these responses by displaying the values across the actuator array and across the focal plane. The actuator responses clearly demonstrates how the pupil and Lyot stop influence the strength of each actuators influence. 

\begin{figure}[H]
    \centering
    \includegraphics[scale=0.25]{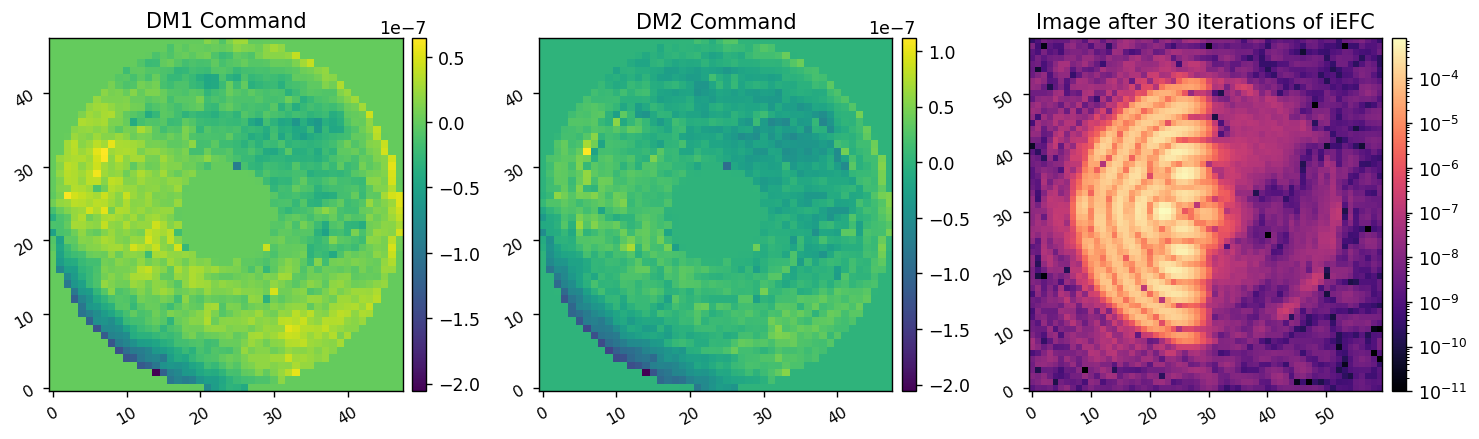}
    \caption{The final dark-hole for the HLC after 30 iterations of iEFC. The final normalized intensity in the control region is about 6.2E-8.}
    \label{fig:hlc-iefc-dark-hole}
\end{figure}

Figure \ref{fig:hlc-iefc-dark-hole} presents the results of iEFC for this imaging mode after 30 iterations. However, iEFC was recalibrated after every 5 iterations to avoid a diverging solution due to the relatively large stroke required for this mode. A weighted least squares regularization was employed when inverting the response matrix to generate the control matrix for this system. The regularization condition was varied between $10^{-1}$ and $10^{-3}$ as well since using the smaller regularization conditions was found to work well for a few number of iterations, but would eventually cause the algorithm to diverge due to larger strokes being used. 

\section{SPC-WFOV Monochromatic Simulations}
\label{sec:spc-wfov-mono-sims}
The SPC-WFOV imaging mode is designed to yield a dark-hole from $6\lambda/D$ to $20\lambda/D$. For this mode, the WFC loop begins with completely flat DMs in both the FALCO and iEFC models. Due to the lack of any initial correction, the normalized intensities are 3.28E-5 and 3.04e-5 for the FALCO and iEFC models respectively. 

\begin{figure}[H]
    \centering
    \raisebox{-0.5\height}{\includegraphics[scale=0.5]{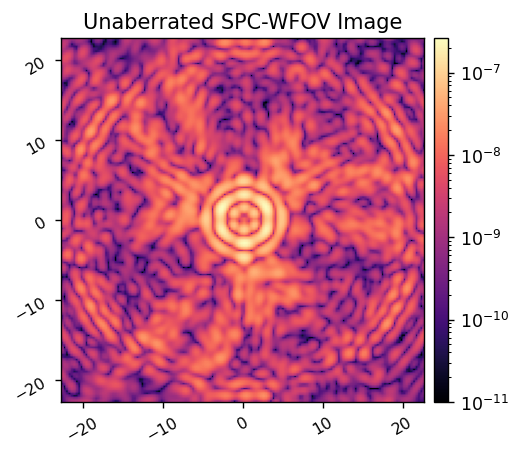}}
    \raisebox{-0.5\height}{\includegraphics[scale=0.5]{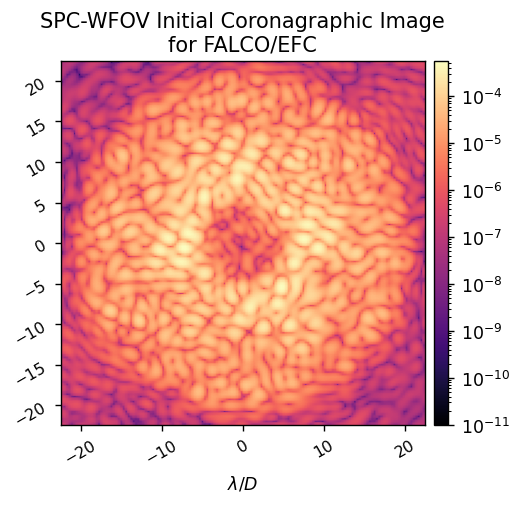}}
    \raisebox{-0.5\height}{\includegraphics[scale=0.5]{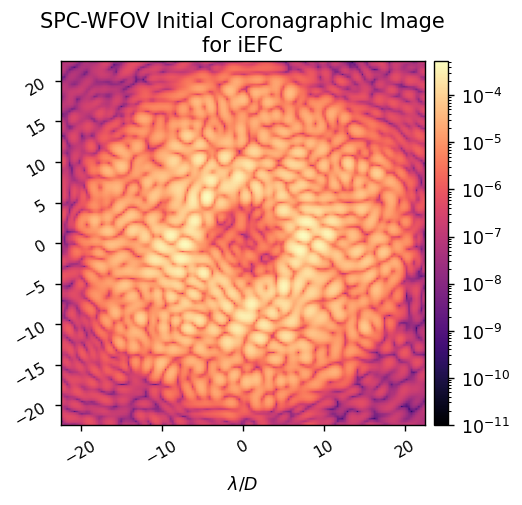}}
    \caption{For this mode, the initial state of the DM actuators are all zero so they are not displayed. The unaberrated image to the left displays what the expected dark-hole for this system should look like, although there is still pupil defocus included in the model of the system resulting in the streaks throughout the desired dark-hole region. The aberrated image illustrates the initial coronagraph state for the FALCO/EFC and POPPY/iEFC simulations.}
    \label{fig:spc-wfov-initial-states}
\end{figure}

For this mode, a full annular control region is defined as the design of the SPC mode makes creating a dark-hole less challenging since an ideal (unaberrated) wavefront would yield a dark-hole as demonstrated in Figure \ref{fig:spc-wfov-initial-states}. Note that the image presented for the unaberrated case still has some pupil defocus included as that is a factor embedded in the optical design of the coronagraph. The HOWFC loop will be able to correct for this pupil defocus as well. 

\subsection{FALCO}
\label{sec:spc-wfov-mono-falco-sims}
\begin{figure}[H]
    \centering
    \raisebox{-0.5\height}{\includegraphics[scale=0.4]{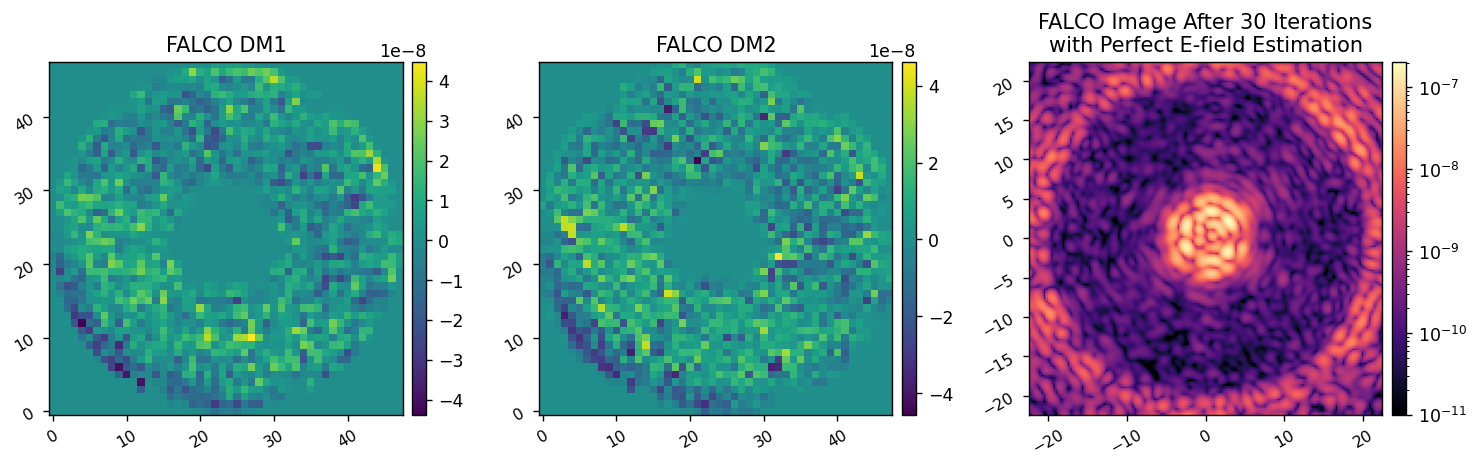}}
    \raisebox{-0.5\height}{\includegraphics[scale=0.3]{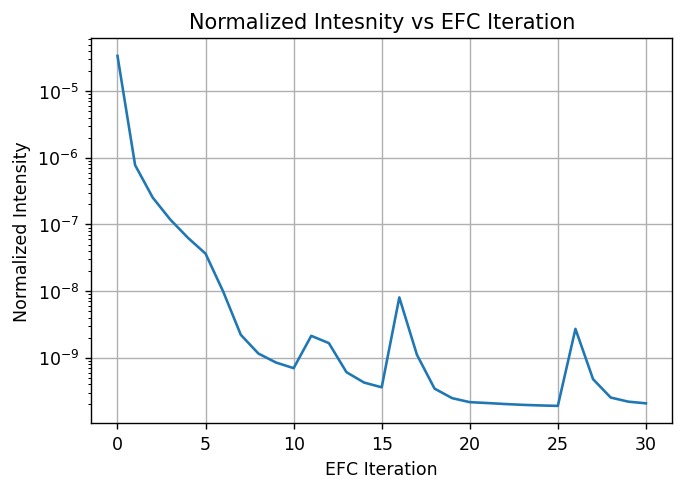}}
    \caption{After 30 iterations of EFC, the DM commands are found to generate a mean contrast of 2.1E-10 within the control region displayed in the right-most image.}
    \label{fig:spc-falco-results}
\end{figure}

EFC is found to yield deep contrast levels within 30 iterations for this imaging mode. While not as critical, relinearization is still performed with FALCO every 5 iterations. Each iteration is using perfect estimation of the electric-field using the end-to-end model. The final DM commands shown in Figure \ref{fig:spc-falco-results} have RMS actuator heights of 10.7nm and 12.3nm respectively. 

The deep contrast attained with this mode is likely due to smaller regularization conditions being used such that each iteration performs a significant amount of correction to improve the contrast. This is only possible given that the Jacobian remains in the linear regime and the estimate of the electric-field is accurate. If the estimate is inaccurate, the DM commands calculated with small regularizations can quickly degrade the contrast by mistakenly generating constructive interference in the focal plane. Given that perfect electric-field estimation is used, each iteration of EFC significantly improves contrast making the algorithm converge to solution within relatively few iterations. 

\subsection{iEFC}
\begin{figure}[H]
    \centering
    \includegraphics[scale=0.5]{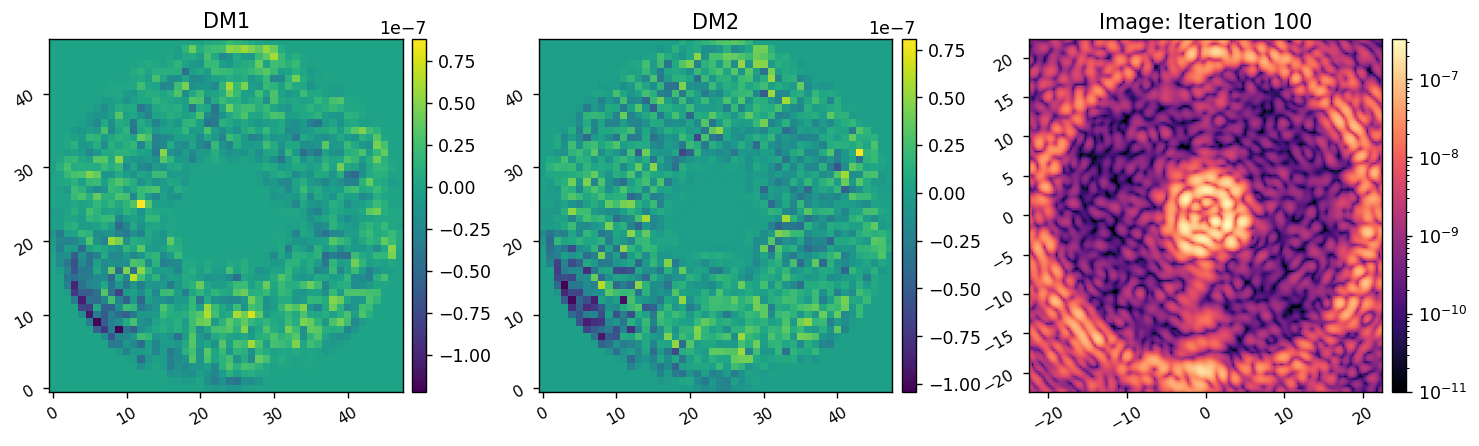}
    \caption{The mean NI is 3.503E-9}
    \label{fig:spc-iefc-dark-hole}
\end{figure}

To perform iEFC with the SPC-WFOV mode, single actuator pokes were calibrated similar what was done for the HLC. However, instead using two Fourier probes, the pair of probes used are also two single actuator pokes. Given that a single actuator poke can be approximated as a small Gaussian in the pupil plane, these pokes generate a much broader response in the focal plane that spans the desired control region. Using these calibration modes and probes, the dark-hole presented in Figure \ref{fig:spc-iefc-dark-hole} is achieved after 100 iterations of iEFC. 

While the dark-hole achieved does not reach the same contrast as the EFC simulation with FALCO, the contrast on the order of 5E-9 is within the desired range for the mode. An important distinction between this mode and the HLC is that no recalibration was performed throughout the 100 iterations. Given that the calibration stage of iEFC can be the most cumbersome if long exposure times are required, this mode is likely more suitable for iEFC than the HLC. 

% \subsection{Demonstration of iEFC with Poke modes on testbed}
% As a proof of concept, the single actuator poke modes are used to calibrate iEFC on the SCOOB testbed at the University of Arizona. This testbed uses a charge 6 VVC and a single 10.2mm Boston Micromachines DM with 952 actuators (hence why an annular dark-hole is not attempted). Figure \ref{fig:scoob-iefc-poke-modes-dark-hole} illustrates the results of using these single actuator pokes to calibrate iEFC. The amplitude used for each poke was 5nm while the amplitude for the Fourier probes was 25nm. 

% \begin{figure}[H]
%     \centering
%     \raisebox{-0.5\height}{\includegraphics[scale=0.3]{figures/iefc_scoob_start.png}}
%     \raisebox{-0.44\height}{\includegraphics[scale=0.25]{figures/iefc_scoob_dm.png}}
%     \raisebox{-0.5\height}{\includegraphics[scale=0.3]{figures/iefc_scoob_dark_hole.png}} % use gist_heat colormap for real testbed data
%     \caption{Demonstrated here is a dark-hole created by calibrating single actuator poke modes with Fourier probes, an example of which was shown earlier in Section \ref{sec:hlc-mono-sims}. To the left is the initial coronagraphic frame from the testbed followed by the final DM command and the final coronagraphic frame after N iterations of iEFC.}
%     \label{fig:scoob-iefc-poke-modes-dark-hole}
% \end{figure}

\section{iEFC with shot noise, read noise, and dark-current}
\begin{figure}[H]
    \centering
    \raisebox{-0.5\height}{\includegraphics[scale=0.19]{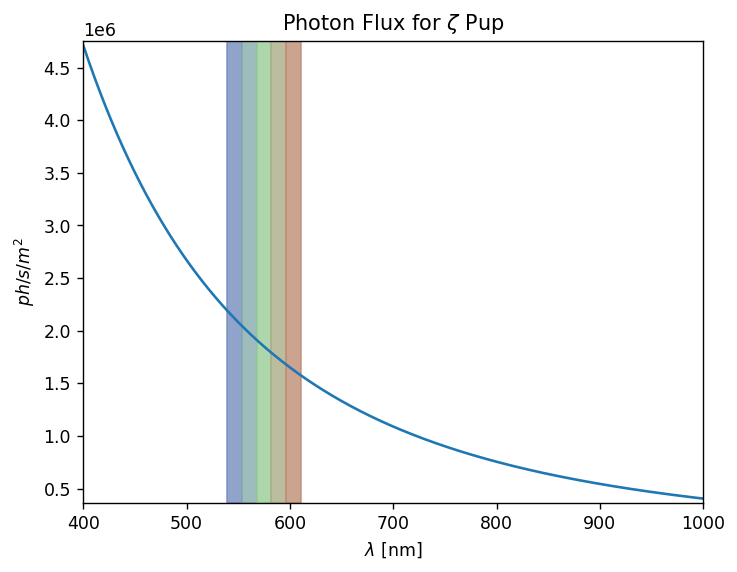}}
    \raisebox{-0.5\height}{\includegraphics[scale=0.23]{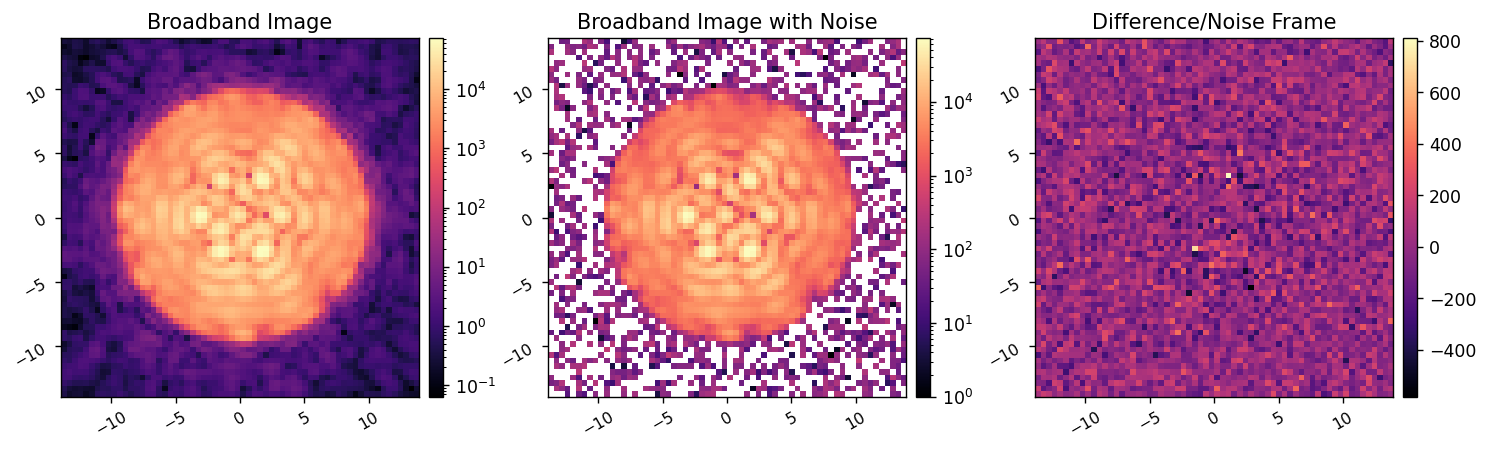}}
    \caption{Example of a broadband image calculated using 5 wavelengths within a 10\% bandpass for the HLC mode. The flux for each wavelength is estimated by integrating the flux curve calculated from the blackbody spectrum for $\zeta$ Puppis over the 5 subdivisions highlighted.}
    \label{fig:broadband-image-example}
\end{figure}

For conducting simulations with noise, broadband images are used for each imaging mode assuming the reference star for wavefront control to be $\zeta$ Puppis. Here, the filters intended for science observations are not simulated since the wider bandwidth of the science filters make WFC algorithms less effective. Instead, a narrower WFC filter is chosen for each imaging mode. These filters are the 2.6$\%$ bandpass centered at 575nm and the 2.9$\%$ bandpass centered at 825nm for the HLC and SPC-WFOV modes respectively\cite{}. Using a simple blackbody model of $\zeta$ Puppis, the flux for three wavelengths within each of the chosen bandpasses are estimated in units of photons/s/m$^2$ using an integration of the blackbody model with three subdivided wavelength regions similar to what is presented in Figure \ref{fig:broadband-image-example}. The equation used to generate the blackbody model for $\zeta$ Puppis is $$B_{\lambda} = \frac{2hc^2}{\lambda^5} \frac{1}{\exp{[\frac{hc}{\lambda k_B T}]} - 1}$$ where $B_\lambda$ is the spectral radiance given in W/m$^2$/sr/nm. The total spectral flux is found by multiplying the spectral radiance by the solid angle of the star estimated by $SA=\pi R_{star}/d$ where $d$ is the distance to the star. The parameters used for the $\zeta$ Pupis are $T=40,000$K\cite{zeta-pup-temp}, $R_{star} = 14R_\odot$\cite{zeta-pup-radius}, and $d=300$pc\cite{zeta-pup-radius}. 

A wavefront for each wavelength is then propagated through the physical optics models with each wavefront being weighted by the estimated flux value and the spatial area of a pixel within the propagated wavefront. Once propagated, an incoherent sum of all three wavefronts is used to simulate the estimated flux in photons/s for every pixel in the image plane. Each broadband image is then converted into individual counts by multiplying the image by a given exposure time. Photon noise applied the image which is then converted to counts by applying a specified gain. Since a bright reference star is being assumed for the WFC stage, the gain values used are within the threshold for the detector to be operated in analog mode (gain$<$1000). The read noise and dark current are then calculated based off the read noise standard deviation and dark current rate which are estimated to be 120$\mathrm{e^-}$/pixel and 0.05$\mathrm{e^-}$/pixel/hour respectively\footnote{\href{https://roman.ipac.caltech.edu/sims/Param_db.html\#coronagraph_det}{https://roman.ipac.caltech.edu/sims/Param\_db.html\#coronagraph\_det}}. Figure \ref{fig:broadband-image-example} includes an example of the broadband spectrum used to calculate the flux for five wavelengths within a $10\%$ band centered at 575nm. A broadband image is also presented with and without noise. For the iEFC simulations, the broadband images are still normalized after noise is applied to remain consistent with the definition of normalized intensity from Section \ref{sec:intro}.

\begin{figure}[H]
    \centering
    \raisebox{-0.5\height}{\includegraphics[scale=0.25]{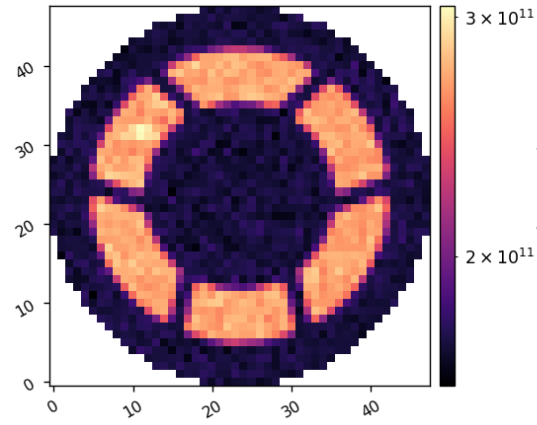}}
    \raisebox{-0.5\height}{\includegraphics[scale=0.25]{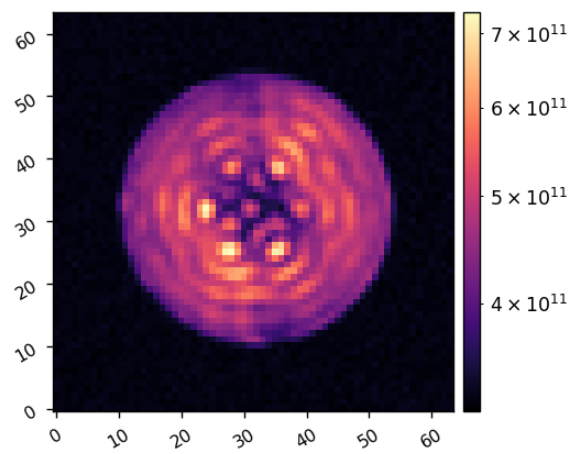}}
    \raisebox{-0.5\height}{\includegraphics[scale=0.3]{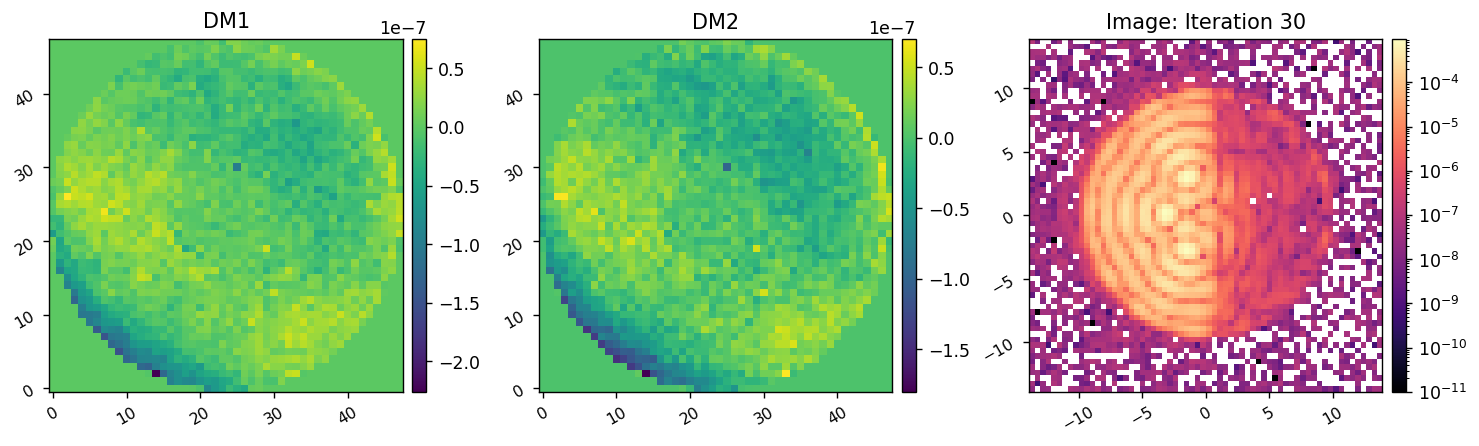}}
    % \raisebox{-0.5\height}{\includegraphics[scale=0.35]{figures/bbspc_wfov_dark_hole_contrast.png}}
    \caption{In the top-left is the RSS response for each actuator of DM1 after calibrating iEFC with noise. The top-right is the RSS response of each pixel in the focal plane. Along the bottom row are the DM commands and the final simulated image of a single wavefront control bandpass after 30 iterations of iEFC. Here, the final mean contrast is 1.3E-6. No recalibration was performed for these iterations of iEFC.}
    \label{fig:bbhlc-noisy-dark-hole}
\end{figure}

The results of the simulation for a single WFC bandpass with the HLC imaging mode is presented in Figure \ref{fig:bbhlc-noisy-dark-hole}. This simulation utilizes 30 iterations of iEFC; however, no recalibrations are performed during this simulation, as the purpose is to estimate how much a single calibration can achieve when noise is included. This is critical because if recalibrations are desired, factors such as gain and exposure time will have to be adjusted for each calibration as the contrast improves. Here, the simulated exposure time is 2s with unity gain for the single calibration. During the 30 iterations simulated, the exposure time is increased to 100s in order to acquire enough SNR to sample the dark-hole. 

\begin{figure}[H]
    \centering
    \raisebox{-0.5\height}{\includegraphics[scale=0.225]{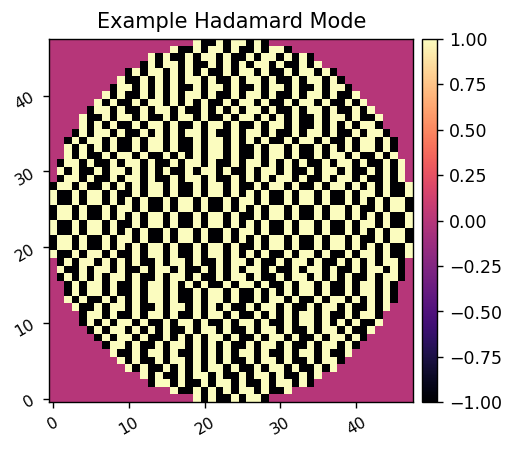}}
    \raisebox{-0.5\height}{\includegraphics[scale=0.225]{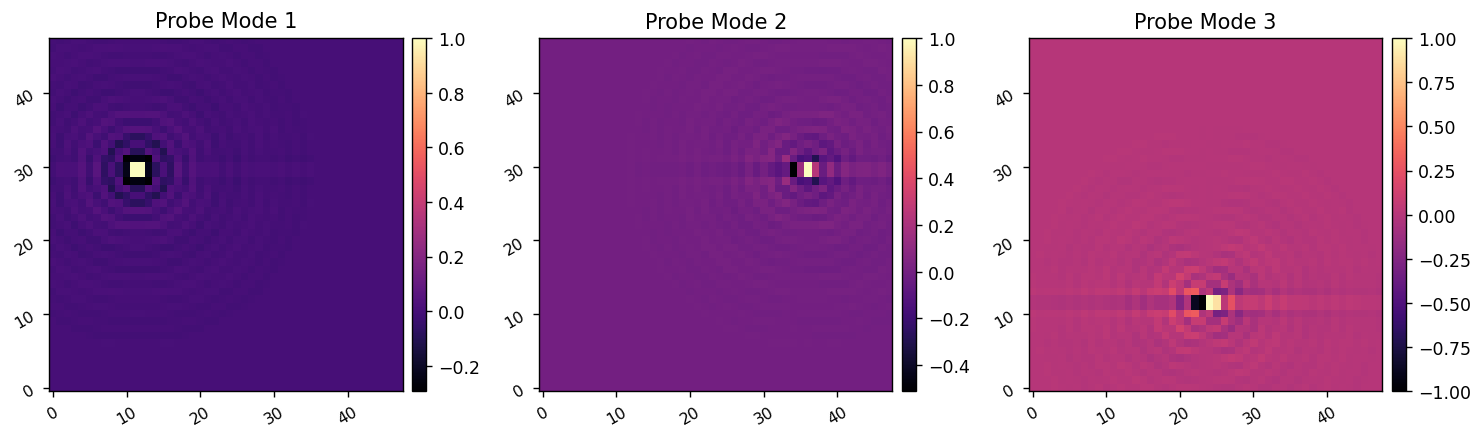}}
    \raisebox{-0.5\height}{\includegraphics[scale=0.3]{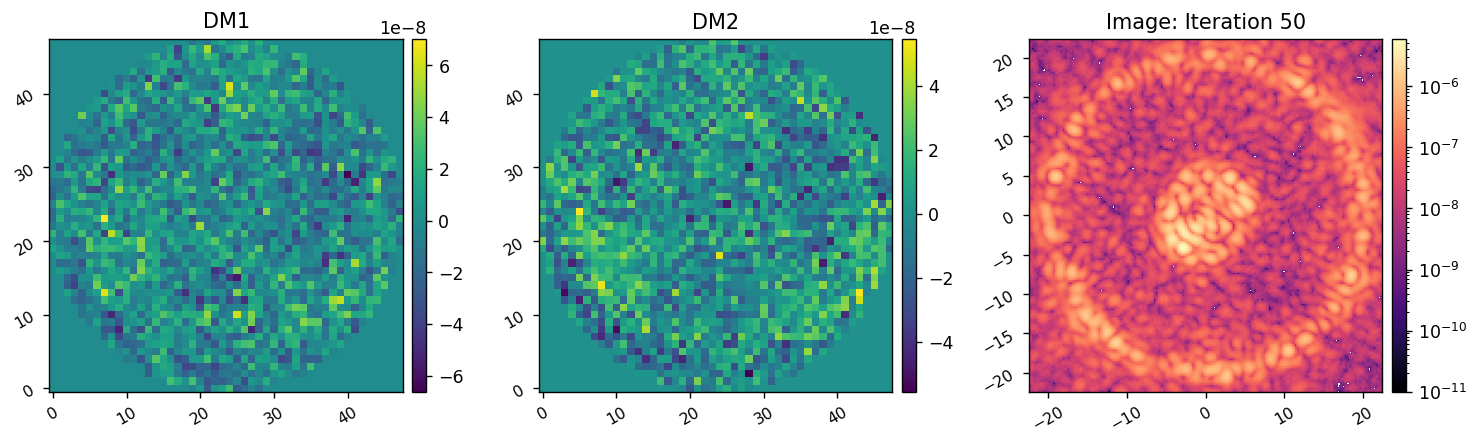}}
    \caption{In the top row is an example Hadamard mode along with the three Fourier probes used to calibrate iEFC for the SPC-WFOV mode. The results of iEFC after 50 iterations are presented in the bottom row. Here, a 2 second exposure time was used to calibrate iEFC while the exposure time was increased over the course of the 50 iterations to 3 minutes. The final mean contrast is 9.0E-8.}
    \label{fig:bbspc-noisy-dark-hole}
\end{figure}

For the SPC-WFOV mode, a two second exposure time was used for the calibration of iEFC. Notably, instead of using single actuator poke modes, a set of 2048 Hadamard modes, an example of which is included in Figure \ref{fig:bbspc-noisy-dark-hole}, was calibrated on each DM (4096 modes total). In addition, three shifted Fourier probes were used instead of two as using three was found to increase the effectiveness of the response matrix. The three probes are shifted in different directions as the actuators that are commanded by the probe mode tend to acquire stronger responses during calibration, which translates to those actuators often using more stroke during the iEFC algorithm. The shift in different directions attempts to mitigate this by distributing the probes across a larger number of actuators.

With the exposure time being two seconds and 4096 total modes being calibrated with three probes, the estimated time to calibrate iEFC would be about 27hrs. Over the course of the 50 iEFC iterations, the exposure time is adjusted as contrast improves. The final exposure time used for the last set of iterations is 3minutes, meaning each iteration will require about 18 minutes of total exposure time. 

\section{Conclusions}
Here, iEFC has been extended to utilize two deformable mirrors and shown to produce an annular dark-hole in simulations. With this capability, iEFC has been applied to both the HLC and SPC-WFOV imaging modes of the Roman Coronagraph. Simulations of the HLC mode demonstrate that iEFC can produce adequate contrast levels within a half dark-hole. While the ideal case of EFC is demonstrated to yield better contrasts, the simulation using PWP to estimate the electric-field demonstrates that the performance of EFC will be highly dependent upon how well the electric-field can be estimated. In this case, iEFC is demonstrated to out perform EFC with the chosen probe and calibration modes if multiple calibrations are used for iEFC. The monochromatic simulations for the SPC-WFOV mode demonstrate that iEFC can be suitable for an annular dark-hole on Roman as well. The SPC-WFOV mode is of particular interest for iEFC given that a single calibration can be used for far more iterations using this mode. Therefore, if the coronagraph is in a state with poor contrast, iEFC can be calibrated with shorter exposure times and used as an initial WFC method until the coronagraph is within a regime where EFC is more applicable. 

The primary concern for iEFC will be the time required for each calibration stage. The current simulations estimate that with approximately 20-30hours of calibration, iEFC can be performed over a single WFC bandpass. In Haffert et al.\cite{haffert-iefc} iEFC is said to be easily extended to broadband imaging by concatenating monochromatic or individual narrow band response matrices into a single response matrix for a wider bandpass. Here, broadband iEFC was only demonstrated for a single WFC bandpass of each imaging mode, but this could be extended to larger bandpasses in the future. The drawbacks for this method are the calibration time will be increased for each narrow band filter response data is required for. Another possibility, if long exposure times are required, could be to use an instrument model to calculate a response matrix. This will avoid the affects of noise in the calibration as well as reduce the calibration time. This idea will make iEFC sensitive to model-based errors similar to EFC, but there is still a potential benefit in that an estimate of the electric-field is still not required, so estimation errors will not propagate to the solution for the DM actuators. An additional factor that requires consideration for future work will be the stability of the telescope during calibration. Overall, iEFC could be a suitable alternative to EFC for initial contrast improvements since deeper contrasts will require longer exposure times making the calibration of iEFC unfeasible. Future telescopes with larger apertures and greater throughput, such as a 6.5m VVC design for the HWO\cite{riggs-vvc-for-offax-aperture,belsten-howfsc-algorithms}, will also be more suitable for iEFC since exposure times can be reduced to enable more rapid calibrations. 

\section{Acknowledgements}
Portions of this research were supported by funding from the Technology Research Initiative Fund (TRIF) of the Arizona Board of Regents and by generous anonymous philanthropic donations to the Steward Observatory of the College of Science at the University of Arizona. 

Additionally, A.J. Riggs, Jaren Ashcraft, Kevin Derby, and Aaron Goldtooth have contributed with helpful discussions throughout the span of this research. 

% \section{Appendix}
% \begin{table}[H]
%     \centering
%     \begin{tabular}{c|c}
%          &  \\
%          & 
%     \end{tabular}
%     \caption{Caption}
%     \label{tab:iefc-vars}
% \end{table}

% References
\nocite{*}
\bibliographystyle{spiebib} % makes bibtex use spiebib.bst
\bibliography{citations} % bibliography data in report.bib

\end{document}